\documentclass{wp}
\usepackage[T1]{fontenc}
\usepackage[utf8]{inputenc}
\usepackage{color}
\usepackage{bm}
\usepackage[title]{appendix}
\usepackage{float}
\usepackage{graphicx}
\usepackage{amssymb}
\usepackage{amsmath}
\usepackage{graphicx}
\usepackage{longtable}
\usepackage{enumerate}
\usepackage{blindtext}
\usepackage{booktabs}
\usepackage{multirow}
\usepackage{pdflscape}
\usepackage{enumitem}
\usepackage{listings}

\colorlet{punct}{red!60!black}
\definecolor{background}{HTML}{EEEEEE}
\definecolor{delim}{RGB}{20,105,176}
\colorlet{numb}{magenta!60!black}

\lstdefinelanguage{json}{
    basicstyle=\normalfont\ttfamily,
    numbers=left,
    numberstyle=\scriptsize,
    stepnumber=1,
    numbersep=8pt,
    showstringspaces=false,
    breaklines=true,
    frame=lines,
    backgroundcolor=\color{background},
    literate=
     *{0}{{{\color{numb}0}}}{1}
      {1}{{{\color{numb}1}}}{1}
      {2}{{{\color{numb}2}}}{1}
      {3}{{{\color{numb}3}}}{1}
      {4}{{{\color{numb}4}}}{1}
      {5}{{{\color{numb}5}}}{1}
      {6}{{{\color{numb}6}}}{1}
      {7}{{{\color{numb}7}}}{1}
      {8}{{{\color{numb}8}}}{1}
      {9}{{{\color{numb}9}}}{1}
      {:}{{{\color{punct}{:}}}}{1}
      {,}{{{\color{punct}{,}}}}{1}
      {\{}{{{\color{delim}{\{}}}}{1}
      {\}}{{{\color{delim}{\}}}}}{1}
      {[}{{{\color{delim}{[}}}}{1}
      {]}{{{\color{delim}{]}}}}{1},
}

\usepackage[natbib=true,backend=biber, style=authoryear, isbn=false, url=true, doi=false, maxbibnames=6, minbibnames=3, maxcitenames=1, firstinits=true,terseinits=true, eprint=false,dashed=false]{biblatex}

\renewbibmacro{in:}{%
  \ifentrytype{article}{}{\printtext{\bibstring{in}\intitlepunct}}}
\renewbibmacro*{volume+number+eid}{%
  \printfield{volume}%
  \printfield{number}%
  :
  \printfield{eid}}

\DeclareFieldFormat{postnote}{#1}
\DeclareFieldFormat{multipostnote}{#1}

\DeclareDelimFormat[bib]{nametitledelim}{\addspace}
% ...
\DeclareFieldFormat[article]{number}{\mkbibparens{#1}}
\DeclareFieldFormat[article]{pages}{#1}
\DeclareFieldFormat[article,misc]{title}{#1}
\DeclareNameAlias{sortname}{family-given}
\DeclareDelimFormat{finalnamedelim}{\addspace\bibstring{and}\space}
\AtEveryBibitem{\clearfield{month}}
\AtEveryBibitem{\clearfield{day}}

\makeatother

\title{From newswire to nexus}
\subtitle{Using text-based actor embeddings and transformer networks to forecast conflict dynamics\thanks{The research was funded by the European Research Council, grant number 101055176-ANTICIPATE, by Riksbanken Jubileumsfond grant number M21-0002 Societies at Risk and by the Research council of Norway through the Uncertainty of Forecasting Fatalities in Armed Conflict (UFFAC) grant number 334977. Computation was carried out on the Alvis clusters of the National Academic Infrastructure for Supercomputing in Sweden, grant number NAISS 2023/5-551. The authors would like to thank Benjamin Radford, Joakim Nivre and Paola Vesco for insightful comments.}}

\begin{document}

\author[1,2]{Mihai Croicu} 
\author[2]{Simon Polichinel von der Maase}
\affil[1]{Department of Peace and Conflict Research, Uppsala University}
\affil[2]{Peace Research Institute, Oslo}


\runningauthor{Croicu and von der Maase}
\shorttitle{From Newswire to Nexus}

\maketitle
\begin{abstract}
This study advances the field of conflict forecasting by using text-based actor embeddings with transformer models to predict dynamic changes in violent conflict patterns at the actor level. More specifically, we combine newswire texts with structured conflict event data and leverage recent advances in Natural Language Processing (NLP) techniques to forecast escalations and de-escalations among conflicting actors, such as governments, militias, separatist movements, and terrorists. This new approach accurately and promptly captures the inherently volatile patterns of violent conflicts, which existing methods have not been able to achieve. To create this framework, we began by curating and annotating a vast international newswire corpus, leveraging hand-labeled event data from the Uppsala Conflict Data Program. By using this hybrid dataset,  our models can incorporate the textual context of news sources along with the precision and detail of structured event data. This combination enables us to make both dynamic and granular predictions about conflict developments. We validate our approach through rigorous back-testing against historical events, demonstrating superior out-of-sample predictive power. We find that our approach is quite effective in identifying and predicting phases of conflict escalation and de-escalation, surpassing the capabilities of traditional models. By focusing on actor interactions, our explicit goal is to provide actionable insights to policymakers, humanitarian organizations, and peacekeeping operations in order to enable targeted and effective intervention strategies.
\end{abstract}

\section{Introduction}

For decades, scholars and practitioners have been engaged in the quest to generate reliable forecasts of political violence, social unrest, regime instability, and civil conflict. At their core, such efforts all seek to generate useful estimates of future conflict risks and future conflict processes, such as onset, escalation, severity, and termination.

To achieve this goal, a variety of methodologies have been applied. If we narrow our focus to the quantitative domain, we can categorize them broadly into three main approaches. The first focuses on socioeconomic and demographic factors to identify quasi-static, or even inert, regions at risk, in e.g. \citet{ward2017lessons}. The second focuses on spatio-temporal conflict histories and past records of battle event data to identify mechanistic spillover and diffusion effects, in e.g. \citet{hegre2021views2020, rod2024review}. The third utilizes changes in the volume and topic of news reporting in order to attempt to leverage low-granularity (usually country-level) signals that a conflict onset is about to take place, in e.g. \citet{mueller2024introducing}.

These approaches work well at identifying 1. "conflict statics" -- i.e. baseline and slow-changing conflict risks, usually at aggregate levels, and 2. at predicting spatio-temporal patterns of conflict once conflict has broken out \citep{hegre2022lessons, hegre2021views2020, mueller2022hard}. However, they are less successful at capturing \textit{conflict dynamics}, specifically predicting fluctuations in conflict intensities. This includes the onset of new conflicts, escalation processes that result in increased intensities, de-escalation processes that reduce conflict activity, and conflict terminations \citep{hegre2022lessons, mueller2022hard, randahl2022predicting}.

To effectively address this challenge, we must shift our predictive focus from the spatio-temporal level to the determinants of these dynamics, examining the individuals carrying out the fighting – the armed actors themselves. These actors, often directly or indirectly referenced in the source material for conflict data (such as news texts), operate within relatively well-defined geographical and temporal boundaries, with a finite set of likely interactions. Furthermore, each actor has unique objectives, histories, strategies, and tactics. Thus, constructing models that can accurately predict the interactions among these actors – and any changes in their interactions – is both a theoretically sound and a practical endeavor.

In this paper, we propose a new approach explicitly aimed at predicting these dynamics. This novel approach combines the precise signals from event data with the rich contextual information from text data, with a dedicated focus on conflicting actors. Specifically, we leverage the fact that the Uppsala Conflict Data Program (UCDP) Georeferenced Event Data (GED) is inherently built around annotations of text that has been manually extracted from Factiva. This results in a richly annotated vast corpus of text data, that we can directly use for forecasting. This further allows us to "back-label" text pieces automatically sourced from Factiva; thus combining detailed structured information on armed actors and conflict events with the comprehensive context found in unstructured text data.\par

We then use this annotated text dataset to re-train a relatively small Large Language Model (LLM) called ConfliBERT \citep{hu2022conflibert}, in order to extend the corpus of UCDP texts on fatal interactions to include richer, more contextual information on the behavior of those dyads.

Next, we take this expanded corpus, which includes both violent outcomes and non-violent actor-level contextual information, and use it to train two larger language models. These models, a (very) large, state-of-the-art, open source 7 billion parameter model -- Mistral 7B \citep{jiang2023mistral}, and a smaller model focused on information extraction -- DeBERTa \citep{he2020deberta}, intend to identify conflict dynamics explicitly, predicting them out of sample with dynamic time windows, conflict escalation, de-escalation, and plateaus at the dyad level.

To limit the corpus size of this paper and focus solely on dynamics, we have chosen to examine the 25 most frequently reported on conflict dyads within the news corpus described below. These were selected from the last six months of the training period (June to December 2021).\footnote{This choice of dyads has some limited implications on evaluation, essentially making evaluation harder than their equivalent on the full set. A further discussion on this subject is made in the web appendix.}

Our approach demonstrates the ability to accurately forecast dynamic developments in violent conflicts, surpassing the predictive capabilities of classical conflict history models. The contribution is threefold: first, utilizing actors to merge the precision of structured event data with the contextual depth of unstructured text data; second, employing near state-of-the-art (SOTA) language models to directly forecast conflict; and third, directly forecasting dynamics and momentum instead of severity of incidence.\footnote{Essentially, forecasting escalatory and de-escalatory dynamics is forecasting the first derivative of the classical severity/incidence measures such as fatality or event counts.} Moreover, we discuss how these innovative methods can address and mitigate some of the most substantial challenges facing current approaches to conflict forecasting.

\section{Literature review and theoretical considerations}
Before delving into our methodology, we first provide a brief overview of existing approaches to conflict forecasting and their respective shortcomings.

\subsection{Existing approaches} 

At its core, conflict forecasting seeks to generate reliable predictions of future conflict events, ideally with high spatio-temporal granularity. These forecasts aim to provide insights into the timing, location, probability, and potential severity of future conflicts, with battle-related fatalities often used as an indicator of the presence and magnitude of violent conflict \citep{perry_2013, chadefaux_2014, mueller_2016, hegre2019views, hegre2021views2020}.\par

Reviewing the landscape of current models reveals three main types. 

The first comprises models based on "structural features" such as socio-economic and demographic factors, health indicators, political regime characteristics, access to natural resources, and terrain features \citep{ward2017lessons, hegre2019views, hegre2021views2020}. These models are adept at identifying regions or countries at risk -- being suitable to identify long-term, near-static trends. However, due to the static nature of these data sources, fall short of accurately predicting the timing or dynamic evolution of conflict, being completely unsuitable to forecast any kind of conflict dynamics.\par


The second approach focuses on models that use past spatio-temporal conflict patterns. Such models employ features like lagged conflict fatalities, frequency of conflict occurrence, elapsed time since the last conflict, and conflicts in adjacent areas \citep{hegre2019views, hegre2021views2020}. Recent advances in these models include the use of deep learning to automatically derive spatio-temporal features, enhancing their conflict forecasting capabilities \citep{malone2022recurrent, radford2022high, maase2022conflictnet}. Currently, these models are, by some margin, the best-performing solutions at our disposal when it comes to making predictions at a highly disaggregated level -- such as within $0.5\times0.5$ decimal degree spatial grid cells commonly used for conflict forecasting \citep{hegre2019views, hegre2021views2020, hegre2022lessons, radford2022high, malone2022recurrent, vesco2022united, maase2022conflictnet}. Yet, since they effectively derive their prediction power from learning past patterns and extrapolating these patterns into the future, their effectiveness in forecasting conflict dynamics is very limited. Due to their mechanics -- past conflict history determine future predictions -- they have very limited ability in capturing escalatory and de-escalatory patterns\footnote{i.e. when conflicts will increase and decrease in intensity.} before the trend is already evident in the historical data. Furthermore, they will have no ability to predict when new conflicts will appear or when existing conflicts will conversely end \citep{mueller2022hard}.

The third approach attempts to address the above-described lack in ability to predict conflict dynamics, by scanning for predictive signal in large newswire corpora, that can (at least theoretically) contain escalatory and de-escalatory signals  \citep{chadefaux_2014, mueller_2016, mueller2022hard, mueller2024introducing, mueller2022using}. Although this approach shows promise in predicting the onset of sporadic events and anticipating escalatory and de-escalatory trends in existing conflicts, it faces difficulties due to its conventional approach. In all cases, these methods have an extremely high-level focus: the authors model not the text itself, but the prevalence and change in the count, number (usually 15–25, a small figure), and proportion of automatically generated topics found in news articles \citep{chadefaux_2014, mueller_2016, mueller2024introducing}. Further, these topics are normally generated in a fully unsupervised manner -- usually using classical topic-modelling tools such as Latent Dirichlet Allocation (LDA) \citep{chadefaux_2014, mueller_2016, mueller2024introducing}, again limiting the signal that is extracted from text. This limits the signals extracted from the texts. Apart from sorting articles into categories and calculating the relative and absolute proportions of each category, the models in this approach disregard the actual text, and thus lose the valuable signals it contains. This limitation often means that these methods are only useful for highly aggregated data, such as at the country level, which is similar to the models in the first approach.

Naturally, combining these approaches can be done to develop more comprehensive, robust, and versatile early warning systems. An example of such a system is the Violence and Impact Early Warning System (VIEWS), a machine-learning ensemble that blends a large roster of diverse models \citep{hegre2019views, hegre2021views2020}. Yet despite combining these three approaches, VIEWS still faces challenges in predicting conflict dynamics, such as escalation and de-escalation patterns as well as conflict onsets and terminations.

However, these conflict dynamics are essential for generating actionable and robust forecasts. In fact, there is a significant demand for models that can provide predictions for rapid developments, especially on a sub-national scale \citep{Caldwell2022}. 

\subsection{Gap in research: Actors and agency}


As already noted, the best predictor for future conflict patterns currently at our disposal is past conflict patterns. Traditionally, this predictive power has been attributed to two closely connected phenomena: \textit{conflict traps} \citep{walter2004does, collier2002understanding, beck1998taking, collier2003breaking, hegre2017evaluating, hegre2021can} and \textit{conflict diffusion} \citep{buhaug2008contagion, ol2010afghanistan, schutte2011diffusion, crost2015conflict, bara_2017}. None of these phenomena, however, are actually (in real life) purely spatio-temporal phenomena. Grid cells do not wage war – it is armed actors, such as states and rebel groups, who engage with each  other, and their own structures and agency determine the outcome and risk of future escalation. This fact has been remarkably well-studied in inferential conflict research, with seminal works dating back to \citet{Kalyvas_2006}, \citet{Cunningham_Gleditsch_Salehyan_2009} or \citet{metternich2013antigovernment}. These researchers all explain the need to explicitly model actor behaviors and interactions, and explicitly determine causal or quasi-causal explanations for conflict dynamics (onset, escalation, de-escalation) in actors’ agency and interplay.

Furthermore, recent studies have demonstrated clear causal links between the agency of actors and spatio-temporal patterns such as diffusion, essentially showing that these spatio-temporal dynamics are to a large extent instruments --- indirect readings -- of actor-level and actor-network level decisions and agency \citep{kim2023spatial}. Essentially, conflict forecasting relies heavily on spatial data, which serves as \emph{a high-variance predictive proxy} for numerous unobserved or poorly measured factors generated by actors or their interactions. Examples include the breakdown of a ceasefire, changing strategic incentives, or increased ethnic tensions \citep{maase2022conflictnet}. Given that these manifest in conflict data with significant attenuation, noise, and with a (varying) time delay, it becomes self-evident that actors should be modelled explicitly in a predictive framework.

To date, the only explicit attempt to incorporate actors and dyads in conflict prediction was made by \citet{metternich2019predicting}. Their research finds that, despite what was expected, this model’s predictive power is actually mixed. In fact, it is either below or at most on par with the current state-of-the-art that only considers conflict history. This is due to two significant hinders: a data issue and a modeling issue.

On the data front, there is very limited up-to-date data on both actors and their activities. Actor-level features such as the Non-State Actors in Armed Conflict Dataset (NSA) \citep{Cunningham_Gleditsch_Salehyan_2009} are both low-resolution and mostly static in nature, conveying a much weaker signal compared to the multitude of country-level and spatial-level indicators available to us if proxying via the spatio-temporal route. Similarly, data on actors' conflict activity is solely defined by the same fatality figures and event count metrics that are available through datasets such as the Uppsala Conflict Data Program (UCDP) Georeferenced Event Dataset \citep{Sundberg_2013, Croicu_Sundberg_2017} or the Armed Conflict Location and Event Dataset (ACLED) \citep{raleigh2010introducing}. These do not provide actor-level features specifically; rather, all the signals they contain are available via the spatio-temporal route.

On the modelling front, developments have been hampered by a difficult methodological problem -- conventional quantitative approaches require assumption of independence that cannot be used with complex interplays of actor behaviors. Attempts have been made to address these, either through the use of graph neural networks \cite{brandt2022conflict} or classical inferential network models \citep{kim2023spatial}, but their inherent estimation complexity combined with the lack of actor-level features made them perform worse than classical approaches.

So, there is still a gap in our understanding. We know that tracking actor agency and behavior, along with changes in these patterns, and modeling them as a prediction task will enable us to capture relevant high-variance signals that are distinct from spatio-temporal conflict patterns. We also know that these will generally increase prediction power, especially in the critical area of conflict dynamics, such as onset, escalation, de-escalation and termination. This is because different actors may use different tactics and strategies, which are not well accounted for in the current models. However, we do not have sufficient data or an established modelling framework to do so.

The next step is to look into the origin of our main conflict data and explore whether we can extract either implicit or explicit actor-level signal from it.

\subsection{Where does the conflict data come from?}

As mentioned, we rely on data from large, human-curated event datasets such as the UCDP GED or ACLED to analyze past conflict patterns (known as conflict history). These datasets attempt to capture detailed information about past and ongoing instances of lethal violence on a global scale, often down to the level of individual villages, and to provide accompanying information on severity, actors, dates, and locations.

But these datasets are not created from thin air. Rather, they are simply secondary data collections – aggregations and distillations of source materials. In fact, most of the source material used to generate these conflict datasets consists of newswire texts produced by global news agencies such as Reuters, Agence France-Presse, BBC Monitoring, and Xinhua. These texts are sourced through an aggregator like LexisNexis or Factiva \citep{Croicu_Sundberg_2017, weidmann2016closer}. This is, indeed, the same source used by most research employing the third forecasting approach -- the analysis of topic proportions in text, which is also utilized by \citet{mueller2022using}. 

Because of the enormous size of the corpora used for data extraction and the entirely manual annotation process, the amount of available data for forecasting is generally limited. Usually just a few features are available, such as location, date, and identifiers for the actors involved, together with some basic predefined classes to which the event can be assigned (for instance, a three-tier categorization of violence) and an estimated level of intensity (such as the number of fatalities) \citep{Croicu_Sundberg_2017}. This severely limits the ability to use dynamic spatio-temporal features in conflict forecasting exercises, both in terms of their nature and level of detail. Further, the process of manual curating the data is time-consuming, and even the earliest candidate-quality signal data is delayed by at least a few weeks from the actual event date \citep{HegreEtAl_2020}.

Essentially, the use of these spatio-temporal conflict patterns involves a two-step forecasting process. First, organizations like UCDP or ACLED manually annotate conflict events, and then forecasters like VIEWS carry out the modeling step.

We thus ask ourselves a question: \textit{Can we use this same corpus to model actor-level behavior explicitly in a forecasting framework?} This can be reduced to two more specific questions: 

\begin{enumerate}
    \item Q1. Can we build a model that directly forecasts from text as input data, without the intermediate step of (manual) data collection, extraction and curation?
    \item Q2. Can we develop a model operating at the actor level with predictive power at that level, thus proving that we can extract actor-level features?
\end{enumerate}

\subsection{Conflict patterns and the potential of LLMs for forecasting}

The past few years have led to an explosion in the development and deployment of \textit{large language models} (LLM). These are extremely large machine-learning models consisting of between tens of million to tens of billion learned parameters. LLMs are designed specifically to learn from natural text as input, and are built in such a way as to make inference in the form of natural text. These models are generally trained on vast corpora of text, with a general language task as the objective function. These functions can take many forms, but examples include predicting a missing word or finishing a sentence \citep{bommasani2021opportunities}.\footnote{This general natural language task, instead of a more specialized objective, make them so-called \textit{foundational models}.}

Typically, LLMs are built around a transformer architecture \citep{attention} -- a neural network architecture structured around the attention mechanism. This mechanism models language in terms of key-query-value patterns between all tokens and all other tokens in the text. Trained on vast corpora of text, such models learn both linguistic knowledge (grammar, word use etc.) and general information about the world at large \citep{bommasani2021opportunities}. While these models can be used for a plethora of objectives (e.g. text generation, classification, regression), a common feature is that text is represented within LLMs as embeddings -- distilled numeric vectors densely encoding linguistic information (i.e. words in their context) \citep{bommasani2021opportunities}.

Two flavors of these models exist: decoder-based models like GPT \citep{achiam2023gpt}, Mistral \citep{jiang2023mistral}, Bloom etc. and encoder-based models such as BERT \citep{he2020deberta} or BART. These models differ in how the attention mechanism is constructed. In decoder-based models, attention is applied in a masked manner -- at each token, the model essentially obscures what comes after the current word (or token) -- thus, the model can only learn from the text that has come before any given word. Conversely, in encoder based approaches, there is no masking of content coming after the current token -- the model has access to the entire piece of text at each token, and can learn by "reading" the text bidirectionally \citep{radford2018improving}.\footnote{Essentially, a decoder model attempts to replicate the actions humans do when reading texts for the first time, whereas an encoder model replicates the process of information extraction from text. Training these models also reflects these -- most decoder-based models are trained on the accuracy of predicting the next word; encoder-based models are usually trained on the accuracy of filling gaps in texts.} These differences make encoder-based approaches more suitable to document classification and information extraction, where text needs to be distilled to non-textual information. Conversely, decoder-based approaches, which explicitly mimic how speech is performed, are better for generative processes like question answering or translation \citep{radford2018improving}.

These can then be adapted to a specific task through one of two techniques, fine-tuning or prompting. Fine-tuning can be defined as modifying the network architecture of the LLM to fit a given task, and continuing the training on new data using the modified architecture. The process is usually done by adding a final layer to the network, depending on the task \citep{bommasani2021opportunities}.\footnote{This last layer can be a logistic regression model (known in this context as a sigmoid activation) if the task is binary classification; it can be a sequential recursive multinomial model for generating new text or answering questions, and it can even be another neural network in e.g. multi-agent or mixture-of-expert approaches. \citep{bommasani2021opportunities}.} Domain-specific datasets, usually different to those used in the original training rounds, are then used to continue the training of the modified model. This continued training starts from the fitted weights of the original model. These weights are then adjusted by optimizing the model for a new specific objective function. This objective function is specific for a given task -- e.g. classifying news articles or segmenting medical imagery. This results in both modified architecture and weights, fit for a highly-specific task. 

Conversely, prompting is supplying further (contextual) information to the LLM as part of its textual input data without any further (re)-training \citep{bommasani2021opportunities}.\footnote{In classical regression notation, where a model is defined as $\hat{y} \leftarrow f(\beta X)$ where $f$ is a fixed machine-learning architecture, $\beta$ is generated by minimizing a loss such that $\beta = argmin(L(y,\hat{y}))$, fine-tuning is providing a new $L_{user}$ and a new series of $y,X$ pairs with the goal of generating a new set of $\beta_{user}$ whereas prompting is providing the best possible fixed text string (called a prompt) as part of the input data such that the combination of prompt and input data (the ${X_{prompt},X}$ pair) provided to the existing model, results in the minimization of the loss function without altering the parameter set $\beta_{model}$.} Essentially, prompting a model is providing instructions and context information to an LLM, and use its recursive generative property to propagate that context into the answers \citep{bommasani2021opportunities}. 

Because fine-tuning changes the trained weights of the model through further training, it requires both the model architecture itself and the pre-trained weights to be available -- i.e. the model being open-sourced -- whereas prompting requires only having the model available for inference. This makes models such as GPT-4, that are not open-source, not open for fine-tuning.\footnote{Some manufacturers make some very limited form of LLM fine-tuning available through their agent creation toolkits. However, in these cases, many of the decisions required for fine-tuning are already made by the manufacturers for a very narrow class of problems very different from forecasting armed conflict.}

LLMs are uniquely suited for our task of analyzing large text corpora because of their inherent ability to directly handle text as an input signal. Further, the ability of LLMs to be adapted to a particular task makes LLMs specifically suited to implicitly identify salient actor-level features and dynamics. Additionally, LLMs learn their own representations from the raw text signal and establish a link between those representations and the target prediction in a single optimization process \citep{bommasani2021opportunities}.\footnote{When such models learn their own representations of features from a complex signal, this is commonly known as \textit{deep learning}.}  So, there is no need to explicitly define what actor-level features we might need or explicitly train models to extract them, since they are latently defined and learned by the model.

Conflict forecasting is a time-series classification process in which the outcomes are time-shifted from the predictors (see e.g. \citet{hegre2021views2020, hegre2019views} for a detailed description of the process). We can therefore use LLMs to allow us to integrate the two steps -- feature extraction and forecasting -- into a single optimization process -- by using the latent implicit feature extraction implicit in such models.\footnote{This is not unique to this task -- in fact, all "deep learning" approaches, from self-driving vehicles to image recognition, as well as all LLM tasks, rely on such implicitly learned and defined features \citep{bommasani2021opportunities}.}

To our knowledge, there are two major systematic reviews that evaluate the performance of LLMs in processing social science texts -- \citet{ziems2024can} and \citet{ollion2023chatgpt}. Both have yielded fairly positive results in the domains of text annotation and information extraction, with the performance of political texts on par with or better than other social science disciplines surveyed. Moreover, for specific tasks like labeling political party affiliation or detecting political misinformation, LLMs approached near-human or even above-human performance. This is further confirmed by \citet{do2022augmented}, who demonstrate that an encoder-based transformer model used to label political texts reaches near-human accuracy, even with limited fine-tuning (few-shot learning), and performs almost as well as a professional research assistant. \citet{gilardi2023chatgpt} and  \citet{gallego} further reinforce these findings on related problems (such as the placement of texts on the right-left political scale).

We therefore have a conceptual answer to our questions above: \textit{train (fine-tune) and use an LLM-based model}.

Within the conflict domain, LLMs have mostly been used for classification purposes, mostly for creation and extraction of dictionaries of noun-verb pattern for further downstream use. Examples include \citet{haffner2023introducing, coped} and \citet{hu2022conflibert}, with the latter providing a pretrained medium-sized LLM specific for conflict contexts. To our knowledge, the only attempt at directly forecasting using an LLM was in \citet{haffner2023introducing}, whose approach was very limited by design. Their LLM-based forecasting model was used as a "strong baseline" to compare against their dictionary-based approach. While their approach used a very small and highly aggregated corpus (yearly conflict reports from the International Crisis Group), and rather coarse target (country-level fatalities) the results obtained were nevertheless very encouraging towards developing a more sophisticated approach.

\subsection{Conflict dynamics, escalation and de-escalation}

To accurately capture the dynamic behavior of actors and predict their behavior, it is essential to set an appropriate target for the models to train on. This is more complicated than it initially seems, mostly because conflict traps are particularly salient, especially in the temporal domain. This results in complex temporal trends, with sudden periods of rapid change followed by long periods of almost complete stationarity.

Previous forecasting work has mostly focused on either incidence, defined as the binary indicator of presence or absence of conflict, \citep{hegre2019views, hegre2021views2020} or severity, defined as the number of observed fatalities or battle events at a fixed point in time \citep{hegre2022forecasting, vesco2022united}. However, models trained on these targets tend, due to the "stickiness" of the process, to quickly regress to the mean -- with optimum performance\footnote{in terms of most conventional metrics.} being achieved by "no change" or "low change" models that predict near constant intensity/incidence \citep{vesco2022united}. Indeed, in most models, up to 90\% of the explanatory power of conflict severity lies in the previous conflict history \citep{hegre2022forecasting, hegre2022lessons, haffner2023introducing, mueller2022hard, maase2022conflictnet} -- making it nearly impossible to train such models to recognize changes and dynamics. 

In the literature, one potential shortcut that has been explored involves the development of custom-built evaluation metrics for conflict dynamics, such as Targeted Absolute Distance with Direction Augmentation (TADDA) \citep{vesco2022united}. However, even such custom-built metrics have been shown to not be sufficient to evaluate change under these extremely stationary trends that classical conflict models exhibit \citep{bracher2023direction}.\footnote{In the case of TADDA, it was shown that it rewarded static, no-change models even more strongly than classical metrics like MSE, in a fashion contrary to initial intentions.\citep{bracher2023direction}.}

To extract dynamics, we cannot use shortcuts but rather need an explicit objective that directly measures the dynamics of escalation and de-escalation, as well as an explicit optimization policy to train a model in a way that captures these dynamics. We thus need to review the existing literature in order to find a solution to this measurement problem.

Conflict dynamics, which refers to the evolution of violence in civil conflicts, has become a crucial and frequently utilized theoretical concept since the publication of \citet{Kalyvas_2006}'s seminal work. (For a comprehensive review of theoretical work in conflict dynamics, see \citet{cederman2017dynamics}). Surprisingly, the development of a measurement model for these dynamics has significantly lagged behind both the theoretical work and micro-level, single-case studies. Therefore, the field lacks a universally accepted standard to measure conflict dynamics, such as escalation and de-escalation, within a conflict.\footnote{Moreover there is some level of inconsistency in how some terms are used: e.g. escalation is sometimes used to only represent the first process that lead to the initial outbreak of a violent conflict from a previously non-violent state (such as protest or deprivation), as in e.g. \citet{gustafson2020hunger}. This is more commonly referred to as "onset" \citep{brosche2023they, mueller2022hard}. For the purpose of this study, we consider escalation to be any dynamic change that increases the intensity of a conflict, and de-escalation to be any dynamic change that decreases the intensity of a conflict -- an onset is always escalation, but not all escalations are onsets.} However, recent studies by \citet{menninga2021battles, randahl2022predicting} and \citet{williams2024bayesian} attempt to address this, leading to the identification of several key features that enable us to establish a common denominator and develop a working definition. 

Most of these approaches treat the dynamics of armed conflict as a succession of temporal "states" defined by a within-state trend and a change point -- i.e. a temporal point where there is a transition between states. Typically, such approaches define three or four such temporal changing points\footnote{E.g. \citet{randahl2022predicting} use "escalation", "de-escalation", "conflict" and "peaceful period" for the four states.}. These usually map to the conventional theoretical underpinnings from e.g. \citet{Kalyvas_2006}. Each such temporal state typically maps directly to an increase or decrease in observed violence over a given temporal window; Each state can be measured directly using solely the recorded battle deaths data \citep{menninga2021battles, randahl2022predicting, williams2024bayesian}. 

For our target, we utilize this theoretically based approach and define four states. These states are labeled as follows: "escalation" (a long-term trend of increased conflict intensity defined in terms of number of fatalities), "reduction" (a long-term trend of decreased conflict intensity), "plateau" (a long-term stable trend) and "zero/no conflict" (close to or zero fatalities).

However, while we agree with previous literature on a working definition, there are significant issues with past measurement approaches, which make their methodological approaches not appropriate. Indeed, all previous attempts resulted in data-driven artifacts, either in terms of spikes or sudden alterations of opposing change points, as noted by \citet{randahl2022predicting}. Even a moving window approach, like that used by \citet{menninga2021battles}, does not eliminate these spikes, which can manifest as immediate successions of escalations and de-escalations instead of a plateau. Proposed solutions, such as using theoretically driven anchors (like ceasefires or talks) to determine window start points or trend direction, are not applicable due to the predictive nature of our task. This is because we do not have access to future events in the test window, as is the case in inferential research.

Instead, we take inspiration from the deep conflict time series decomposition approach introduced by \citet{maase2022currents}. This entails using long-term Gaussian processes to model, and thus determine, these states. This method, unlike the one proposed by \citet{williams2024bayesian}, is explicitly interpretable and computationally cheaper.









\section{Methodology}

\begin{figure}[!p]
    \centering
    \includegraphics[width=1.5\textwidth,angle=90,origin=c]{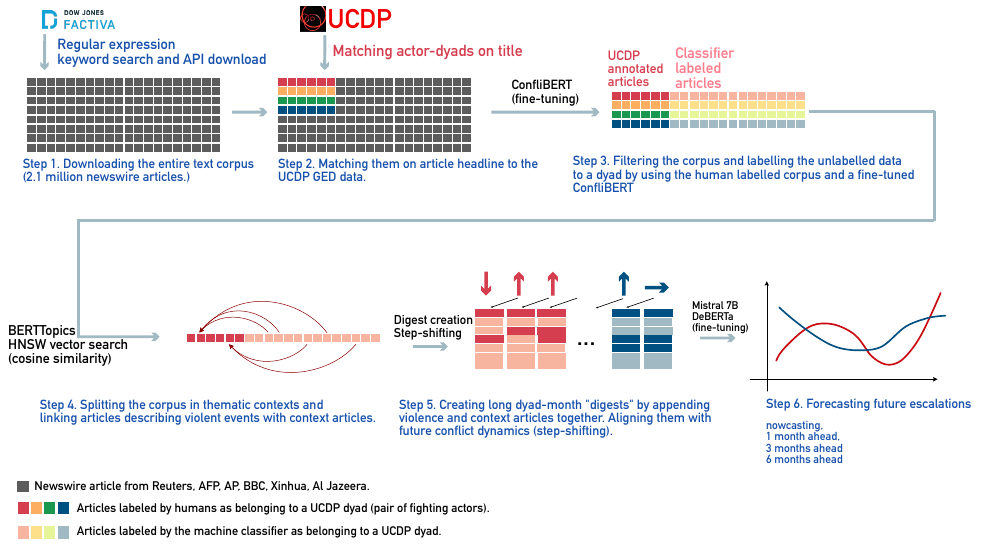}
    \caption{The general architecture of the proposed text-to-forecast end-to-end system. Note that matching with UCDP articles is only needed at training time; the trained classifier can be used for dyad-months not containing any labelled data.}
    \label{fig:aarch}
\end{figure}

After examining existing approaches and their respective challenges, we will now detail our methodology, outlining each step in-turn. Our approach is summarized in Figure \ref{fig:aarch} and consists of five steps: 
\begin{enumerate}
    \item Extracting the unlabeled text corpus;
    \item Partially labeling the text corpus using UCDP GED labeled articles;
    \item Using the partially labeled data in order to train a classifier to filter and label the remainder of the corpus;
    \item Grouping the articles by context and aggregating them based on similarity to create long and relevant dyad-monthly text digests for training and evaluation;
    \item Defining and measuring the target (conflict escalation), and attaching future escalation values to the digest for training and evaluation. This is done by using the direct multi-step forecasting approach that is referred to in the field as step-shifting \citep{hegre2021views2020};
    \item Training and evaluating the final predictive model.
\end{enumerate}

We will describe each of these intermediate steps and present detailed intermediate
evaluations as necessary below.

\subsection{Extracting the corpus}


Our approach relies on two principal sources of data: the same newswire text corpus that is extracted from Factiva, and then scrutinized and manually annotated by UCDP \citep{Factiva2022} and the annotated UCDP GED event data \citep{Croicu_Sundberg_2017}. The UCDP GED records instances of organized violence across the globe, with a focus on fatality counts. It includes precise details on the timing and location of each incident and the actors involved. Notably, this information is derived from Factiva using a systematic dictionary approach to search for potentially relevant articles that can subsequently be manually processed and annotated. Therefore, the UCDP GED also incorporates metadata, such as article headlines, from the source. The UCDP GED covers events from 1989 to 2023, with monthly updates for ongoing incidents provided by the UCDP GED Candidate dataset \citep{HegreEtAl_2020}. These factors make the UCDP a central data source for obtaining detailed, high-resolution training data while also allowing support for the development of "live" near-real-time applications.\par

To extract the news corpus, we use the Factiva API \citep{Factiva2022} and mirror UCDP’s approach used to extract their corpus of articles for manual annotation. We follow their approach in order to generate training data for our modeling, which involves reversing the sequence of operations carried out by UCDP. This method allows us to use UCDP GED events to label Factiva articles, instead of deriving event annotations from the articles. To do this, we use the same query that UCDP uses, as described in \citep{Croicu_Sundberg_2017}:\par

\centerline{\{kill*, die*, injur*, dead*, wound*, massacre*\}\footnote{For replication, see the web appendix for the exact API call we used.}}

To account for the change in how source links are presented by the UCDP in their GED datasets, as well as to balance the need for a large comprehensive dataset against our computational resource constraints, we limited our search to articles published between January 1, 2015, and April 1, 2023.  This approach yielded approximately 2.16 million articles containing one or more of the search (stem) words. To avoid contamination, we divided the data longitudinally in a training set (until the end of 2021) and a test set (starting January 2022).\footnote{As stated before, since we are specifically interested in conflict dynamics, due to API access limitations, we limit ourselves to forecasting the 25 most intensely reported on UCDP dyads in the last six months of the training period -- June--December 2021. A list of these dyads is provided in Appendix A1.}

\par

\subsection{Matching with UCDP GED}


The next step is to match the headlines of all these articles to the headlines connected to the events cataloged in the UCDP GED. Effectively, this pairs relevant articles with a conflict event, allowing the UCDP GED data to act as metadata for the article. It is essential to consider that the actors annotated in the UCDP GED events are not always explicitly mentioned in the articles. Instead, they have been inferred by experienced annotators and area experts working at UCDP who have access to context far beyond what is presented in the individual news piece. Therefore, by re-connecting newswire articles with the UCDP GED data we are effectively enhancing these articles with expert annotations on the actors involved. This pairing process resulted in 54,669 articles that could be matched with a UCDP GED event, leaving around 2.1 million articles that were unmatched by any event yet contained the search string.

These matched events are now "fully labeled" -- i.e. contain those newswire articles that were used explicitly by UCDP coders to extract conflict actor information. Therefore, each of these articles now contain reference to at least one pair of conflict actors -- in UCDP parlance, a dyad \citep{Croicu_Sundberg_2017}.

The rest of the articles that could not be matched to UCDP events can be categorized into two groups conceptually. On one hand, there are articles that refer to a specific conflict dyad, either directly or indirectly, but do not refer to a fatal battle event (we will call these "contexts"). These articles may provide efficient signals for escalatory practices. On the other hand, there are articles that are completely irrelevant and have nothing to do with either political violence or conflict actors. These can refer to topics like sports, corporate news, legal action, or disease\footnote{A large part of the corpus is COVID-related, due to the very dense news coverage of the 2020 pandemic.} -- texts that use the language of armed conflict but involve other human activity.

This problem of irrelevant reporting is widely recognized in the field, and its high prevalence only amplifies the background noise in what is an already weak signal. This problem is not novel, and has been acknowledged since the earliest attempts to extract even data using natural language processing -- e.g. \citet{schrodt2012automated}. While many solutions have been proposed -- e.g. \citet{croicu2015improving} -- we need a novel approach that can do two things -- not only eliminate irrelevant articles, as in previous approaches, but also assign the remaining context articles to dyads in a relevant way.

\subsection{Assigning articles to actor-dyads and filtering out irrelevant items}

\begin{figure}[!htb]
    \centering
    \includegraphics[width=0.80\textwidth]{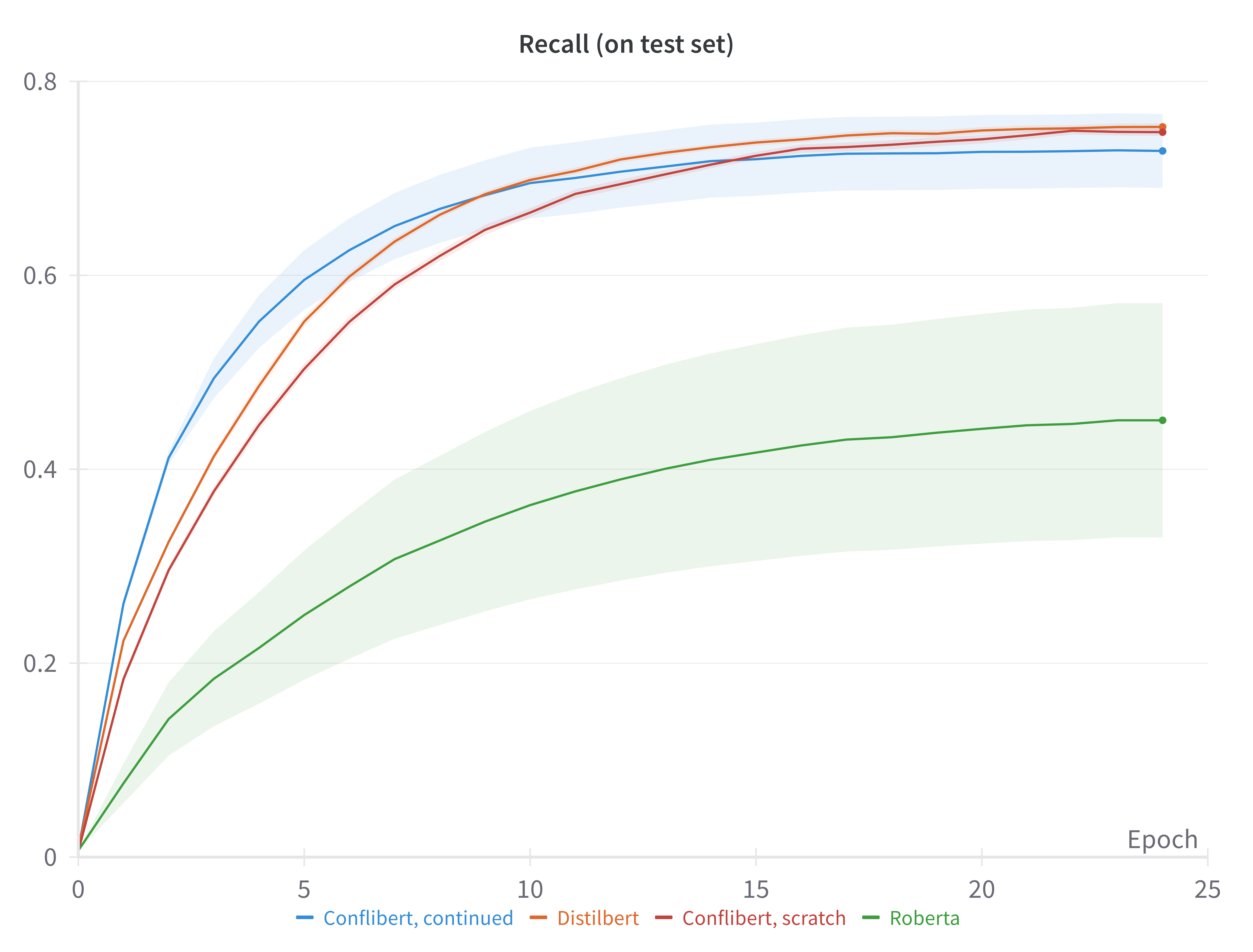}
    \caption{Multi-class micro-aggregated recall plotted against training epoch across the 40 experiments using the four different base BERT-based classifiers. }
    \label{fig:recall}
\end{figure}

As a result of the previous step, however, we still have over 2 million unlabeled articles. Many of these likely refer to the dyads themselves, but not necessarily to battles resulting in fatalities. Instead, they provide vital context into negotiations, strategy, leadership, and other relevant information. This step intends to extract articles that mention a dyad and filter out those that are completely irrelevant. This step follows a straightforward approach to semi-supervised transfer learning as described by \citet{torrey2010transfer}, which involves using a partially annotated corpus to annotate the entire dataset. The choice of a filtering model for irrelevant articles is driven by computational challenges. Because there are a vast number of articles to filter, we need a model that is relatively simple, easy to train, and fast.

We thus need to make one core assumption -- if an article is about a pair of fighting actors, then this information will be found in the introductory section of the article (known as head and lead paragraphs in news terminology). Thus, for performance, we set the tokenizer to truncate any article longer than 256 tokens, since performance of transformer models at inference is $O(n^2)$ with the number of tokens \citep{attention}.\footnote{Most large language models use the token as their main input and output. This is either one word (if it can be found in the fixed-length vocabulary predefined by each model), or a part of a word or even a single letter if not. Vocabularies are predefined, and contain a fixed number of words based on their descending frequencies in the training corpora. Vocabularies are defined such that any word can be represented by a token or series of tokens, even if they were not encountered before. BERT has a vocabulary of 30,000 tokens, Mistral has a vocabulary of 32,768 tokens, GPT-2 (the last open-source version) has a vocabulary of 50,257 tokens \citep{hu2022conflibert, he2020deberta, jiang2023mistral}.}

We then fine-tune a battery of 40 BERT models to predict, for each article, the name of the UCDP dyad to which the article refers.\footnote{The target is all 150 dyads that have at least 25 articles in the annotated corpus, plus a final category, "other". An alternative to this would have been to train a binary classifier for each dyad -- however, this would have been both computationally inefficient and substantively problematic -- as the purpose is to bin events into mutually exclusive classes, which binary classifiers would not have provided.} The reason we train so many models is to be able to evaluate them as a battery\footnote{individual models differ on standard hyperparameters -- learning rates, learning schedules, batch sizes, training samples, etc.} -- instead of selecting a model a priori, we train batteries of 10 ConfliBERT-scratch, 10 ConfliBERT-continued\footnote{these are specific models, pretrained for conflict-aware news corpora.}, 10 of the very light DistilBERT model and 10 of the slightly more complex general purpose RoBERTa model \citep{hu2022conflibert, liao2021improved}. We train ten models in order to be able to avoid overfitting on a single test set as well as the even more problematic model collapse, common with fine-tuning BERT type models. Fine-tuning is carried out and evaluated on a standard 80/20 train/test split of the labelled corpus employing random partitioning.\footnote{One random partition per each model in the battery of 10 models. This entails 43,735 articles in each train set, 10,933 in the test set. Hyperparameter tuning is carried out on one model from each family, following the same approach as in \cite{hu2022conflibert} with a goal to optimize recall.}

Micro-averaged recall\footnote{Recall is defined as the proportion of articles correctly assigned to the correct dyad as a proportion out of all true articles belonging to the class. Micro-averaging involves computing recalls class by class and averaging these at the end; this is particularly useful with unbalanced datasets such as this one, where dyad intensity and newsworthiness is highly variable.} against the amount of training (in epochs)\footnote{An epoch is a pass over the entire training set.} is presented in Figure \ref{fig:recall}.

There is essentially no difference between the performance of models based on ConfliBERT-scratch and DistilBERT (recall 0.7530 vs 0.7478)\footnote{On other metrics, such as accuracy and precision, the two classes of models perform nearly identical}, with ConfliBERT-continued performing slightly worse and RoBERTa substantially worse. 

Since performance for the two best classes of models is excellent and highly consistent, we select the best performing model by recall, as we believe ensembling would not provide any advantage worth the complexity tradeoff. This model is a ConfliBERT-scratch model\footnote{Trained for 30 epochs, learning rate $5 \times 10^{-5}$, Accuracy: 0.6149, F1: 0.5689, Precision: 0.4822, Recall: 0.7976. All relevant scores computed using micro-averaging.} is kept and used to annotate the entire unlabeled corpus.\footnote{Essentially, this model functions similarly to a UCDP coder by assigning newly encountered articles to dyads and providing probabilities for the article to belong to a given dyad.}

We then eliminate any article that has a less than .8 probability of belonging to any of the 25 pre-selected dyads. This leaves us with 273,750 news articles, now completely labeled with one of the 25 dyads. This gives us a fully supervised, labeled corpus while simultaneously providing a reduction of almost 87\% from the original.

\subsection{Generating contextual data}

We are interested in predicting escalation at the dyad-month level, and not at the article level.\footnote{This is for self-evident reasons: a single news article, from a single day, describing a single event or happening is unlikely to ever contain sufficient information to reliably decide whether the whole conflict process is escalating or de-escalating. A very violent event in an otherwise de-escalating conflict will be (correctly) labelled as escalation even though it is a singular event.}

Such aggregation from article to model could be done at two levels:
\begin{enumerate}
    \item Text level -- basically creating human legible "digests" by combining snippets sourced from multiple articles. Each resulting digest will be a long (thousands of words) text, combining pieces from multiple articles in the dyad-month -- essentially a long summary of all the articles in the dyad-month, describing what is happening with a given dyad in a given month.
    \item Embedding level -- by aggregating model-processed numeric vectors containing representations of text) \citep{ziems2024can} using linear algebra techniques. 
\end{enumerate}

We prefer the former for several reasons. The first has to do with transparency and in terpretability. The architecture of the whole approach is already complex, so being able to manually spot-check each step is valuable for avoiding errors. Second, aggregating embeddings would be beneficial if we could optimize it jointly with the base model; unfortunately, this approach is not computationally feasible. Third, the models used for filtering are rough and were selected for computational speed and not discriminatory performance. Therefore, we decide to pursue the first approach of creating digests for each dyad-month. This involves selecting relevant pieces of articles and compiling them into (long) texts that represent the actor-level activity of a whole dyad-month. We then use these digests as the basic unit for both training and forecasting.

However, two challenges remain -- one, the amount of labeled articles for each dyad-month is still very large, with large amounts of redundancies, duplication, and irrelevant content; and two, extracted context articles outweigh articles directly describing violence by as much as 11 to 1. Simple random sampling is unlikely to produce much more than noise. To mitigate this problem, we take inspiration from two sources: \citet{mueller2024introducing}'s topic-proportion models described above and Retrieval Augmented Generation (RAG) \citep{lewis2020retrieval}. RAG -- a recent development in machine-learning -- is a technique to extract and collate contextually similar information to a given piece of text \citep{lewis2020retrieval}. This allows contextually similar text pieces to be collated into a larger and richer piece. RAG exploits the nature of LLM-generated embeddings as high-dimensional vectorial spaces that embed information in their dimensions. Through the use of fast, geometric similarity calculation routines, these allow for easy and extremely fast similarity searches at both linguistic and informational level.

We combine the two approaches in the following way: using the embeddings produced by the ConfliBERT model at the previous step, we fit a deep topic model for each dyad-month using BERTTopic's embedding-based topic model \citep{grootendorst2022bertopic}. We diverge from \citep{mueller2024introducing} not just in fitting dyad-specific models, but also in that we allow the number of topics to vary with the number of articles assigned to each dyad, so that the smallest topic for each dyad is set at 200 articles. This results in between 3 and 21 topics per dyad. We use these topics in two ways:

\paragraph{The low-context approach}

The restricted approach is a simple, classic topic-model based approach to identify the context surrounding the fighting and combine this context with the text on fighting. Essentially, for each dyad-month, this approach extracts the mean textual context (non-fighting behavior) for each topic of behavior the dyad actors have been involved in that month.

We do this by picking the closest five articles to each fitted topic centroid for each month. We then take snippets from these selected articles\footnote{We define a snippet as the first 256 tokens of the head-and-lead paragraph of an article.} and concatenate them into one large piece of text containing snippets from all topics. 

Finally, a digest is created by taking all the snippets from UCDP violent events for a given dyad month and appending the centroid-derived contextual snippets described above, creating a long text describing the entire dyad-month. We refer to this large text as the "low-context digest".

\paragraph{The high-context, RAG approach}

For this, larger, approach, we make full use of RAG. Instead of looking at topic centroids, we try to extract the non-violent context of fighting for each individual event level. We do this by augmenting each event text with contexts relevant for that individual event drawn from each topic.

For each UCDP event, we do this by sampling the closest article in each non-violent topic to that event's source text. We do this search by calculating cosine similarity in embedding space between articles using the hierarchical navigable small world (HNSW) approach described in \cite{foster2023computational} -- a fairly conventional RAG approach. We combine the snippet from the violent event with a snippet from this nearest contextual article for each topic (resulting in 4--22 snippets), creating, for each event, a combined text containing a description of the violent event augmented, via RAG, with close-in-proximity contextual information drawn from nearby topics.

We then take these event-plus-context texts and concatenate them into a large digest for each dyad-month. If the text becomes longer than the context window of the model, we take random samples of texts up to the context limit (creating multiple such digests for each dyad-month). We refer to this large text as the "high-context (RAG) monthly digest".

\subsection{Measuring escalation}

\begin{figure}[!htbp]
    \centering
    \includegraphics[width=\textwidth]{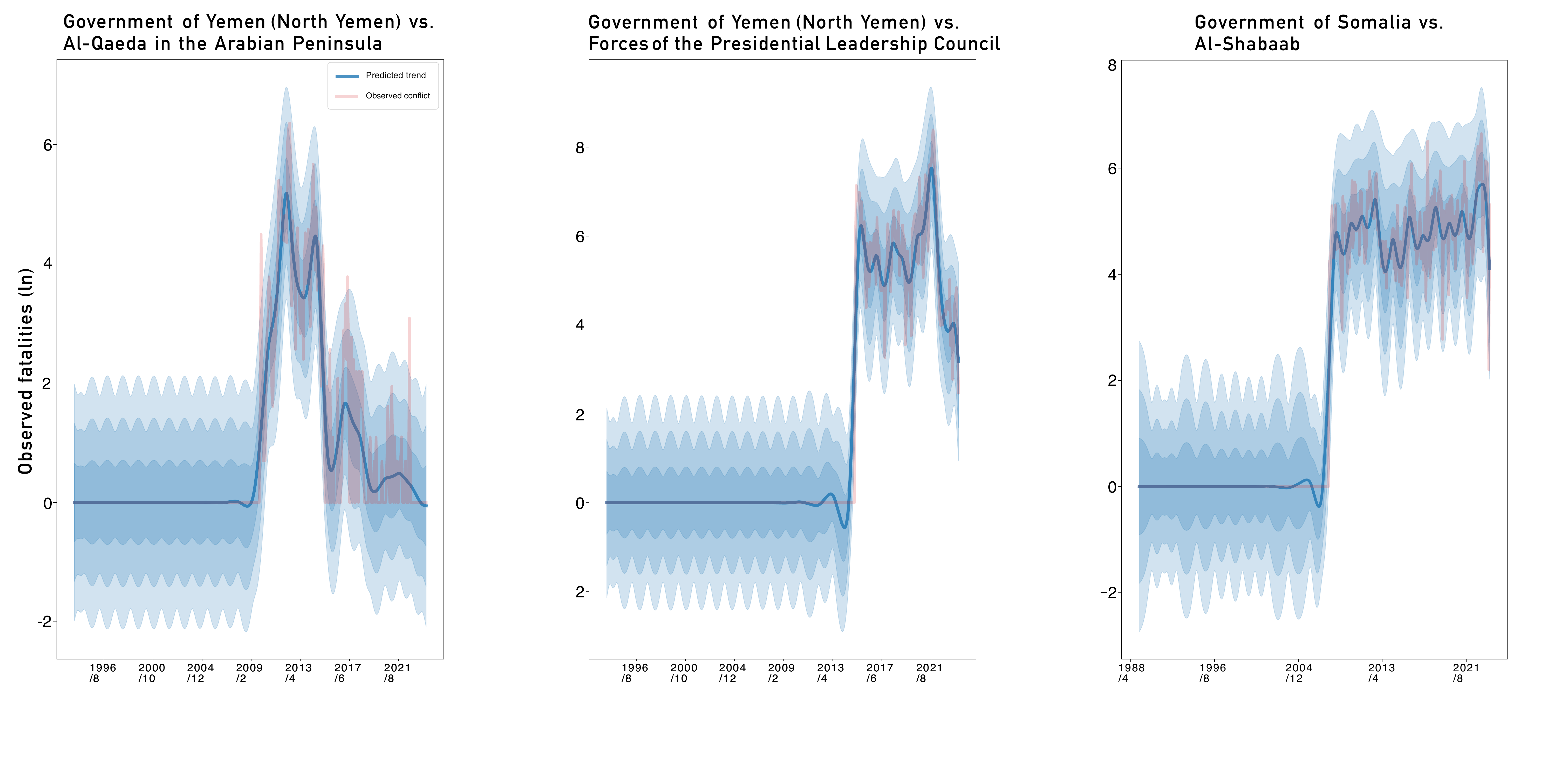}
    \caption{The fitted multi-level temporal Gaussian process smoother for three UCDP dyads in Yemen and Somalia (blue), including 90,95 and 99\% credible intervals against the raw observed data. The first derivative of this trend is used as the target in our forecasting experiment.}
    \label{fig:gpp}
\end{figure}

The next step is measuring the four temporal states (escalation, deescalation, plateau, and zero/no conflict activity) that we defined theoretically above. 

We adapt the solution provided by \citet{maase2022currents}, where the number of log-fatalities ($y_t$) observed in a dyad-month is modeled as a Gaussian process. In this approach, the data is assumed to come from a distribution made up of an infinite series of Gaussian functions. This distribution is characterized by two key parameters: a mean function $\mu$, usually set to 0, and a covariance function $K$. The covariance function determines how any observed point on the temporal axis ($x_t$) is allowed to influence any other point on the same axis as a function of the temporal distance between them \citep{williams2006gaussian}:

\[
y_t =  f(x) + \epsilon \tag{1}
\]

\[
f(x) \sim \mathcal{GP}(\mu=0,K(x_t,x_{t'}))
\tag{2}
\]

We use the Matérn $\frac{3}{2}$ covariate function \citep{williams2006gaussian}, which is one of the standard covariate functions used to model Gaussian processes.  As a partially derivable function, Matérn $\frac{3}{2}$ has been shown by \cite{maase2022currents} to be most effective for temporal modelling of conflict trends:

\[
K_{Matern\frac{3}{2}}(x_t,x_{t'}) = \eta^2 \left(1+ \frac{\sqrt{3|x_t-x_{t'}|^2}}{\ell} \right) exp\left(-\frac{\sqrt{3|x_t-x_{t'}|^2}}{\ell} \right) \tag{3} \label{eq:k_mat}
\]

where $\ell$ (length scale) is the main parameter to be estimated in the model, the distance (in months) at which points influence each other.

We then estimate a hierarchical Gaussian process $f_{dyad}$ for each UCDP dyad, with the observed points being monthly aggregates of the natural logarithm of observed fatality figures in that dyad.  We depart slightly from \cite{maase2022currents}'s approach here, by enabling not only time-points within each dyad to inform each other, but also different dyadic processes ($f_{dyad}$) in the same country to inform each other during the estimation process in a hierarchical model.

The estimation is then carried out using a sparse Bayesian estimation routine based on \cite{amosidentifying}'s approach. For each dyad the estimation is started from a global highly informative prior drawn from \cite{maase2022currents}'s findings -- a LogNormal centered around the main temporal long-term finding in that paper, the estimated global maximum a posteriori (MAP) long-term $\ell$ of 122.38. Results from this estimation are presented graphically in Figure \ref{fig:gpp}. 

However, this estimation will only provide us with smoothed and extracted long-term conflict trends in log-space, $\hat{f}_{dyad}$. To extract escalation states, we take the (numerical) first derivative of the fitted function $\hat{f}_{dyad}$ and discretize it into a four states $S(x_t)$ for each month $x_t$ such that:

\begin{equation}
    S(x_t, {dyad}) = 
    \begin{cases}
      0, \text{or "Peace" if  } y_{t,dyad} = 0 \\
      1, \text{or "Escalation" if  } \hat{f'}_{dyad}(x_t) > \tau \\
      2, \text{or "Plateau" if  } \tau \geq \hat{f'}_{dyad}(x_t) 	\geq -\tau \\
      3, \text{or "De-escalation" if  } \hat{f'}_{dyad}(x_t) < -\tau
    \end{cases}\,.
\tag{4} \label{eq:disc}
\end{equation}

where $\tau$ is a threshold value set at 0.25 (to account for any potential estimation wobbliness) and $f'(x)$ is the first derivative of $f(x)$.

To prevent data leakage from the training to the validation partition, we estimate two Gaussian processes -- one $\hat{f'}_{train, dyad}$ trained on data until the end of the training window (December 2022) and one used for validation $\hat{f'}_{val, dyad}$ until the end of the data (March 2023).\footnote{Due to computational requirements, only these two windows were estimated, allowing us a single, fixed, training and validation window.}

These discretized momentum function will serve as both the training and validation targets for the model. Since they are slopes to the smoothed observed fatality trends, they eliminate the base effect, since $S(x)$ describes observed changes. Traditionally, base effects, i.e. past fatality figures account for up to 90\% of the explanatory power of classical models -- with this approach such base effects are fully accounted for and discounted. Similarly, since $S(x)$ is based on smoothed, decomposed, long-time trends, the spikiness that rolling averages would provide is also discounted.

\subsection{A (deep) LLM-based forecasting method}

Since these monthly digests are lengthy, spanning thousands of words, models with larger context windows are necessary to effectively parse the text.\footnote{Most LLMs have a limited context window: a limited amount of text on which the attention mechanism can generate weights for their key, query, and value weights. Thus, there are hard limits on the length of text a model can "understand", and thus on the length of a sequence. In the case of ConfliBERT this is as low as 512 tokens.}

Instead, we rely on two near-state-of-the art foundational models: an encoder-only model, DeBERTa-v3-large from the BERT family \citep{he2020deberta} and a decoder-based generative model, Mistral 7B v0.3 \citep{jiang2023mistral}. Both offer a much larger parameter space (304 million for DeBERTa and 7 billion for Mistral). Moreover, and more importantly, their different architectures allow them to handle substantially longer texts by processing word-level embeddings differently.\footnote{Mistral can handle nearly infinitely long sequences due to the rotary embedding approach it employs \citep{jiang2023mistral}. whereas DeBERTa can handle up to 24,528 tokens at once through its relative embedding encoding and sharing mechanism \citep{he2020deberta}. In practice, both methods can handle very long texts -- with the available GPU RAM being exhausted before architectural limits are reached.}

In both cases, we set up our forecasting experiment in the same way as the state-of-the-art in violence forecasting approaches (such as VIEWS). That is, we use the approach referred to in conflict forecasting as "step-shifting" \citep{hegre2021views2020, hegre2019views}, or as direct multi-step forecasting in econometric literature \citep{chevillon2007direct}. Therefore, we train models against a target shifted forward $t$ time steps into the future, enabling them to learn the signal so that they can predict $m$ time steps into the future, given data at the forecast horizon.

We set this up in the following manner: we add a softmax (multinomial logistic) classification layer to each LLM taking the last layer of the network as inputs and outputting probabilities for the four classes of escalation (peace, escalation, plateau, deescalation). We then fine-tune each LLM. We do this three times -- once for DeBERTa using the restricted set of digests, once for DeBERTa using the full (including context), and once for Mistral using the full digests.\footnote{We do not fine-tune Mistral on the restricted digest, as computational intensity of training is extremely high -- and the biggest advantage of Mistral is its ability to digest very large and complex texts.} We fine-tune the models using the standard multi-label cross-entropy loss weighted for class imbalance in the train dataset. We fine-tune (continue training) the entire model, including the final classification layer, as is standard.

For DeBERTa, we start by fine-tuning from the open-source weights provided by Microsoft \citep{he2020deberta} via the Huggingface platform. For Mistral, due to its extreme size, we apply two standard optimizations. First, we apply 4-bit quantization, instead of training at the full 32-bit floating point precision, we train at a smaller precision, defined as 4-bit (32) steps carefully chosen to reflect the distribution of the parameters for the layer \citep{hu2023peft}. This allows the model to fit on the largest GPU we had available, an Nvidia A100 with 40 GB of RAM. The second adaptation involves training on a low-rank adaptation of the original model (LORA) \citep{dettmers2024qlora,hu2021lora}, where only the attention heads and the classification head weights are trained; the rest kept frozen at their original states. Both adaptations are quite standard, and have been evaluated in multiple contexts, including some in political science, and to not affect performance \citep{ziems2024can}.\footnote{Hyperparameter tuning in terms of the learning rate, dropoff and number of training epochs was carried out for DeBERTA, but were impossible for Mistral. This was done on a separate evaluation set, with a small randomly sampled dataset from the test window, to reduce computational costs.}

Still, the step-shifting approach requires training a model for each time step. Since training is extremely computationally intensive and requires high-performance GPUs\footnote{The models require the use of high-performance computing clusters. Fine-tuning a single Mistral model, even with LORA and quantization, requires approximately 48 hours on four separate high-performance Nvidia A100s with 40 GB of RAM; whereas fine-tuning DeBERTa requires 14--24 hours on an A40 GPU. At the time of writing, these are top of the line high-performance server GPUs.}, and since we assume the escalatory signal that the text contains is short-term, we only train four time steps for each model. These include nowcasting (detecting escalation in the current month of texts), 1-month ahead, 3-months ahead, and 6-months ahead, resulting in 12 models trained in total.\footnote{At a total cost of over 2,100 GPU-hours, including hyper-parameter tuning.}

\section{Results, forecasting stage}

\begin{figure}[ht!]
    \centering
    \hspace*{-0.94cm} 
    \includegraphics[width=1.15\textwidth]{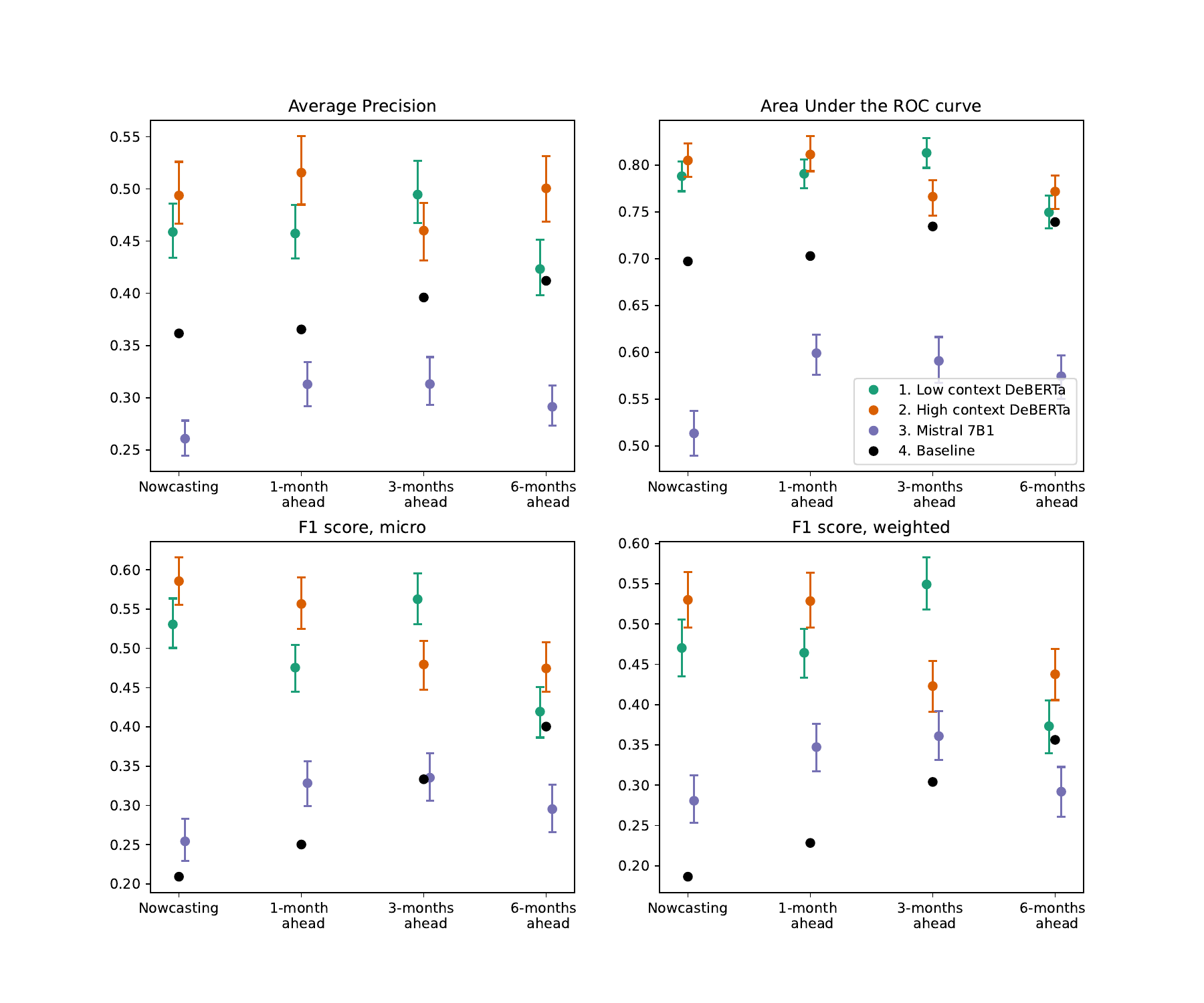}
    \caption{Main metrics for the experiments over four time steps (now-casting, 1-month ahead, 3-months ahead and 6-months ahead). Multi-class prediction (4 classes, escalation, reduction, zero, plateau), micro-aggregation where probabilities are involved (ROC, AP). Confidence intervals were obtained from 1000 bootstraps of the predictions.}
    \label{fig:results}
\end{figure}

Figure \ref{fig:results} shows the main metrics in the 15-months evaluation window, at digest-dyad-month level.\footnote{The total number of observations ("monthly-dyad text digests") is 1,000 instead of the expected 1,125. This is the result of there being some cases where it was impossible to sample more than a single unique digest for each dyad-month, as only a single set of contextual information could be extracted. As a precaution and to keep the entire process comparable, we make sure that in all three cases we evaluate on the exact same test-set structure, with the exact same 1,000 observations dyad-month structure. This allows us to compare between models and methods.} For all aggregate metrics, micro-averaging was chosen, because it better reflects the unbalanced nature of the dataset and the problem we are trying to solve best \citep{grandini2020metrics}. This is a result of the selection process -- using only the most reported 25 dyads and -- leading to the most interesting classes (escalation and de-escalation) being the majority classes. Further, correctly predicting escalation and de-escalation is more interesting than hitting the zeroes and plateaus, which are, in essence, easier to predict. Macro-averaging, the other alternative, would overly emphasize the ability to predict the easy to predict minority classes (zeroes and plateaus), leading to misleading results.

We choose a particularly strong baseline inspired by the \textit{conflictology} benchmark in \citet{Hegre2024A} -- for each dyad we bootstrap the last 12 months of the observed values in the training window (minus the step size to avoid contamination) and aggregate these bootstraps to create a distribution of pseudo-probabilities of these four steps. We then use this empirical distribution as our baseline. This approach is inspired from climatological literature, where similar strong baselines are used for evaluating momentum-driven models. In fact, the benchmark is so strong most models in state-of-the-art prediction competitions of political violence are unable to beat it \citep{Hegre2024A}.

The two DeBERTa-based models perform substantially above the baseline across all metrics for now-casting\footnote{which, unlike the case of conflict history models, can still be highly useful, as the trained model does not require any human annotations.} as well as for one-month ahead forecasts, with average precision (AP, also known as the area under the precision-recall curve) and F1 scores (both weighted and micro-aggregated) being almost twice above the relatively strong baseline chosen.\footnote{We cannot compare our efforts with other approaches such as VIEWS or ACLED Cast, as we focus on the harder task of predicting momentum, and not incidence.} This demonstrates that newswire texts contain significant signal beyond the one provided by past conflict dynamics alone. Further, this shows that models trained solely on text can extract them.

The results are attenuated somewhat but still useful when predicting three months ahead. They yield somewhat more unstable and weaker results, yet are still significantly above baseline performance across all metrics. Additionally, model training for 3-step ahead models was much more prone to collapse, with over 40\% of the RAG-augmented models collapsing during training. This required lowering of the learning rate to extremely low values as well as substantial tweaking of hyper-parameters to avoid collapse.

At six months into the future, the two DeBERTa models almost collapse. They are just slightly better in AP and F1 than the baseline, and are at baseline for AUROC. This is consistent with theoretical expectations. Although newswire reports do contain signal about current events with future impacts and information about immediate future events, such as elections, the further one goes into the future, the less signal there is and the more noise there will be. There are two plausible explanations for this: first, the prevalence of short news cycles, and second, the inherent uncertainty about the future, which often leads to speculation or punditry in news coverage. For a more detailed discussion, see, e.g. \citet{domingo2011vol}. This result represents the overall temporal frontier at which the model is capable to make accurate forecasts, which aligns with previous findings by \citet{cederman2017predicting} and other.

Despite having many more parameters and a more complex architecture, the Mistral model performs substantially worse than the baseline and is essentially useless for forecasting at any step, unlike the DeBERTa models. We cannot provide a conclusive explanation for this because the models involved are highly complex. The complexity of these kinds of LLMs make any traditional interpretation-based debugging approach unfeasible  \citep{bai2021attentions, jain2019attention}. Similarly, ablating the model for debugging purposes was not possible given the significant computational resources required, which greatly exceed those at our disposal.\footnote{Prompting-based approaches on the base Mistral model have been attempted as an alternative approach to fine-tuning. These gave similarly poor results on forecasting tasks. Further, similarly poor results were observed even with simpler tasks such as extracting the number of battle-related fatalities from a text.} 

We did keep the result in the paper, largely to support \citet{ollion2023chatgpt}'s argument that zero-shot and few-shot decoder-only models, both commercial and open-source, require careful handling and rigorous testing. Further, even with such care taken, results may not as favorable as in \citet{gilardi2023chatgpt}'s extremely favorable use-case. Indeed, in our case, even with substantial (many-shot) fine-tuning and a dedicated classification layer, it performed much worse not only than a simpler encoder-based LLM model, but much worse than even a strong baseline based on sampling from the past empirical distribution of conflict.

We will now turn to comparing the two models. Appending conflict-contextual articles, and making a larger digest through RAG techniques, does help predictive performance; the extended model performs better than the simpler model in three out of four cases.\footnote{At three months ahead, the extended, RAG-based high-context model is quite unstable, and requires some changes of starting values in order to not collapse. With substantial increase in epochs (to 25) and decrease of learning rate to $5 \times 10^{-7}$ the extended model does outperform the simpler model. However, this increases training time almost sevenfold. For comparability across steps, we chose, however, to present comparable models.}

However, the difference in performance is not particularly large, and this can be attributed to the nature of the corpus we used. It was repurposed from a corpus of texts related to lethal fighting events that was developed specifically to optimize manual information extraction of these events. Indeed, considering the intended use of the corpus, it is highly likely a lot of signals about non-lethal events (statements from group leaders, observed tactical changes, non-lethal acts of violence, negotiations and agreements, actions of rebel governance, etc.), which we know exist in newswire text \citep{croicu2022reporting}, are being disregarded during the initial stages of the keyword search process.

This is also confirmed when looking at inter-class predictive performance. If we binarize the classes in a "one-versus-rest" approach -- AUPR/AP for predicting escalation is 0.6043 at 1-month ahead and 0.6104 at 3-months ahead for the small model, whereas for predicting de-escalation (versus the rest) it is substantially weaker, at 0.4534 at 1-month ahead and 0.4104 at 3-months ahead. Additionally, at 6-months-ahead, escalation drives the collapse of the models, and exhibits the biggest drop in predictive performance (binary AUPR/AP dropping to 0.4347). Conversely, performance for de-escalation only decreases marginally to 0.4087, indicating that the model is not equally predictive for both classes.\footnote{Prediction for plateau and zero-classes is comparably much more constant than de-escalation. However, we choose not to explicitly discuss these in comparative terms, due to their relative scarcity compared to escalation and de-escalation (this was by design as we selected our experiments on the most intensive dyads).} This is not due to class imbalance -- escalation and reduction are fairly balanced classes, accounting for 51.05\% respectively 32.53\% in the test partition for 1-month ahead forecasting.\footnote{Full balance tables are provided in Tables 3 and 4 of Section 4 in the appendix.}


What could account for this differential in predictive power between escalation and de- escalation? The most likely explanation is the corpus used to train the forecasting model. We began with a corpus specifically extracted based on keywords optimized for recalling high-violence, lethal battle events. This project aimed to minimize human annotation labor and maximize the number of articles in the corpus that contained references to battles and fatalities. Since this baseline corpus is skewed towards reports of violence (increases in fighting), and filters out reports of peaceful activities (negotiations, power-sharing, etc.). Due to this, the classifier itself may exhibit bias towards better predictive performance of escalatory patterns and worse predictive performance of de-escalatory trends.

Given this, it is highly likely that the first step in improving this model and achieving a fully viable attempt at direct modeling of escalation from text involves moving away from a corpus based on repurposing UCDP's GED event corpus. Instead, a vaster corpus of actor activities can be extracted by directly searching for known dyad and actor names. However, due to current licensing restrictions on newswire texts and API access, we are unable to test this hypothesis by extracting a corpus that utilizes dynamic querying of dyad and actor names.

\begin{figure}[ht!]
    \centering
    \hspace*{-1.1cm} 
    \includegraphics[width=1.2\textwidth]{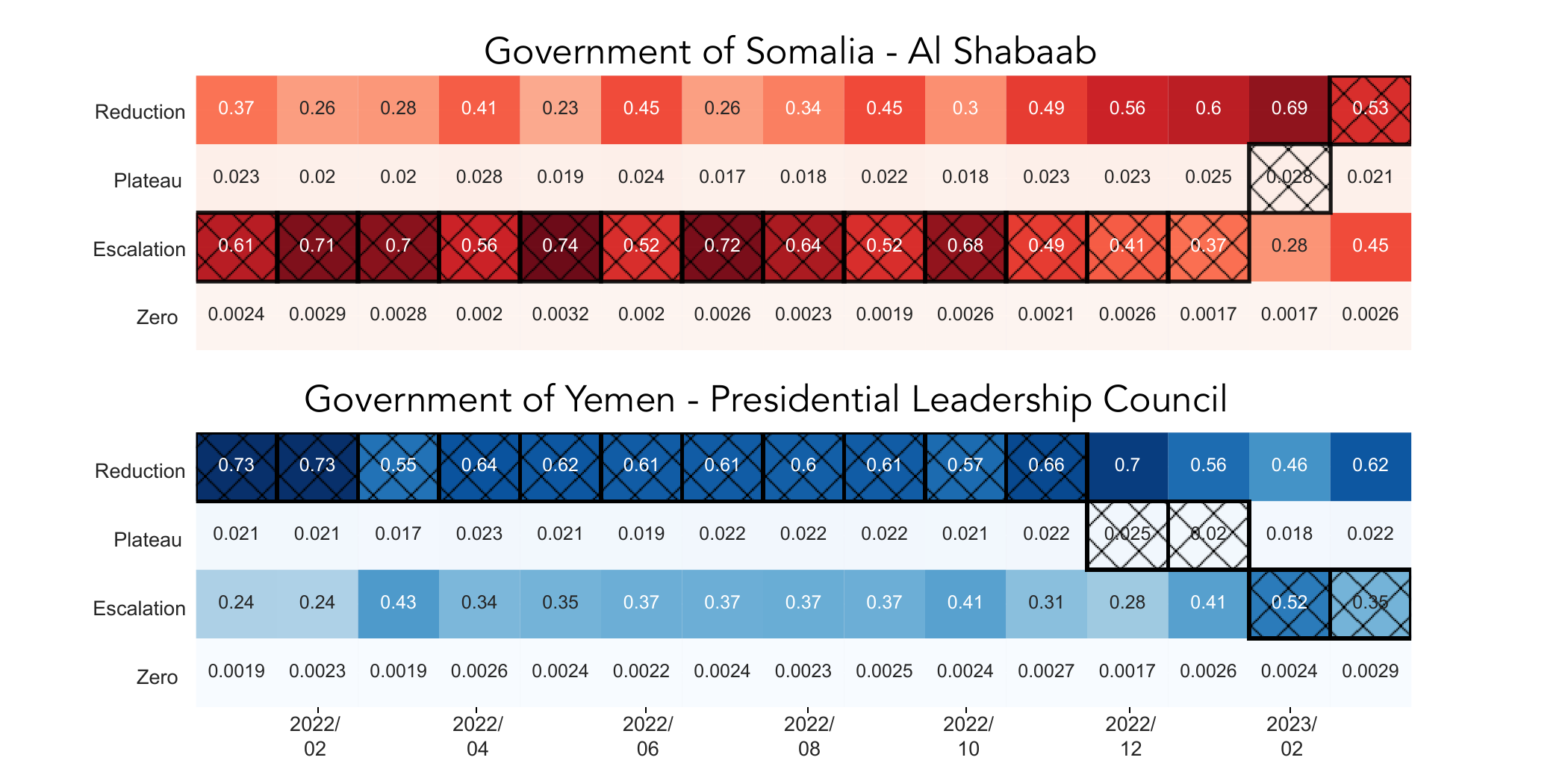}
    \caption{Per class predicted probabilities for two of the dyads containing the most text, forecast one month ahead. Those squares that are outlined in black and covered in hatches represent the actual class, as observed. For both predicted probabilities and displayed actuals, the newswire text used in the model to predict is one month into the past from the forecasting horizon; the calendar month is on the horizontal axis.}
    \label{fig:two_dyads}
\end{figure}

Finally, we turn to specific predictions. Figure \ref{fig:two_dyads} graphically displays the forecasts generated for two of the highest reported dyads in the test set, one that transitions from escalation to de-escalation and the other that transitions in the other direction. In both cases, the large RAG model is used, and both predictions and actuals are one month into the future with regards to the texts used to forecast them. Performance is adequate for both of them, since the models are able to predict state transitions. It is interesting to note that the text signal seems to switch faster and earlier than what our long-term Gaussian process escalation trends would suggest.

\section{Conclusions and future work}

In this paper, we have demonstrated conclusively that newswire texts contain signals that can be used to effectively predict the momentum of armed conflicts at the fighting actor-dyad level over periods of up to six months. By analyzing newswire texts, we can anticipate signals of escalation and de-escalation before they happen. This advancement enables the field to take a step forward in predicting the dynamics of armed conflict. Instead of relying solely on binary incidence and fatality data that are subject to the influence of conflict traps and diffusions, and have significant static base-effects, we can now anticipate future risks of escalation or de-escalation.

Additionally, we have demonstrated through the use of fine-tuned large language models that it is possible to create forecasting models without the intermediate step of expensive and time-consuming manual extraction, distillation, and curation of battle-event information, at least at prediction time. This allows for predicting escalation and de-escalation at three forecasting horizons: nowcasting, one- and three-months ahead -- that were previously deemed unfeasible. 

However, much work remains to be done. Despite a growing body of literature suggesting that they perform well even in zero-shot approaches, the poor performance of state-of-the-art decoder-based models compared to much smaller encoder-based models remains a methodological challenge. 

Moreover, the considerable difference in predictive performance between forecasting escalation and forecasting de-escalation is in itself a complex issue that will necessitate significant efforts in improving corpora extraction and information distillation. This, however, is made very difficult by the extreme cost and usage-rights restrictions in sourcing newswire corpora.

Further, our approach to augmenting violent events with context information, while state-of-the-art, did not explore beyond relatively simple RAG approaches -- mostly as RAG was not previously employed in conflict prediction. Based on the findings, it is clear that while such approaches do provide improvements and are worth carrying out, solutions closer to the theoretical machine-learning state of the art are worth investigating.

Finally, our current approach functions as a stack of separate pipeline components that are optimized individually according to different objectives at each step of the process. Although using LLMs is a common approach in state-of-the-art application development patterns, to the extent that advanced tooling (such as \texttt{Langchain}) now exists to handle this programming pattern \citep{topsakal2023creating}, it is possible and can be worthwhile to explore a more tightly integrated end-to-end architecture. This could perhaps take the shape of a mixture-of-experts ensemble or a reinforcement-learning trained ensemble with an end-to-end optimization policy. However, at this time, given the computational resources available, such an approach is substantially beyond our realm of possibility.

\clearpage
\newpage

\printbibliography

\end{document}